\documentclass[proceedings]{JHEP3}

\PrHEP{PrHEP hep2001}                   
\conference{International Europhysics Conference on HEP}                

\usepackage{epsfig}                   

\title{Study of neutrino induced upgoing muon energy}
\author{\speaker{Eugenio Scapparone} for the MACRO Collaboration\thanks{see
\cite{macro}for the complete author list}\\  
INFN-LNGS, SS 17bis km18+910 61070 Assergi(AQ) Italy\\   
E-mail: \email{Eugenio.Scapparone@lngs.infn.it}}    

\abstract{An estimate of the energy of neutrino-induced muons in MACRO is
provided by a multiple Coulomb scattering measurement. The MACRO original
upward-muon data sample has been subdivided according to the reconstructed
muon energy. The results are interpreted in terms of neutrino oscillation}
\begin{document}
\section{Introduction}
MACRO\cite{Ahlen} can be used as a neutrino detector by measuring 
neutrino induced muon events. 
From the study of the upgoing muon deficit and from the distortion of the 
relative angular
distribution, MACRO provided evidence
for neutrino oscillations\cite{macro}. 
The oscillation probability depends on the ratio $L_{\nu}/E_{\nu}$, 
where $L_{\nu}$ is the distance travelled by neutrinos inside the earth and
$E_{\nu}$ is the neutrino energy: an estimate of this ratio is fundamental for any 
oscillation analysis.
For high energy muons $L_{\nu}$ is properly measured using the 
reconstructed zenith angle of the tracked muon.
As far as the $E_{\nu}$ is concerned, part of the neutrino energy is
carried out by the hadronic component produced 
in the rock below the detector while
the energy carried out by the muon is degraded in the propagation up to the
detector level. Nevertheless, Monte Carlo simulations show that the 
muon energy at the detector level still preserves 
information about the original neutrino energy.

Since MACRO is not equipped with a magnet, the only way to infer the muon 
energy is through the multiple Coulomb scattering (MCS) of muons in the
$\simeq 25$ radiation lengths ($X^{o}$) of detector. 
We use the streamer tube 
system\cite{Ahlen}, which provides the muon
coordinates on a projected view. The other complementary view of the
tracking system (``strip'' view) cannot be used for this purpose since
the space resolution is too poor. 
In MACRO, a muon crossing the whole apparatus has 
$X/X^{o}~\simeq~25$/cos$\theta$ and $y~\simeq$~480/cos$\theta$ cm, 
giving, on the vertical, $\sigma_{x}^{MS}$~$\simeq$~10~cm/E(GeV).
The muon energy estimate can be performed up to a saturation point, 
occurring when $\sigma_{x}^{MS}$ is comparable with the detector space
resolution $\sigma_{x}$. The MACRO streamer tube system, with a cross
section of ($3 \times 3$)~cm$^{2}$, provides $\sigma_{x}$$\simeq$1 cm: the muon energy 
estimate through MCS is possible up to $\simeq$ 10 GeV/$\sqrt{cos\theta}$.

A first energy estimate has been presented in\cite{Bakari},
where the feasibility of this approach was shown. 
The deflection of the muons inside the detector depends on the muon energy
and was measured using the digital information of the limited streamer tube
system. 
The measured event rate vs. $L_\nu/E_\nu$ is in good
agreement with the expectations, assuming neutrino oscillations with 
$\Delta m^{2}$=2.5$\times$$10^{-3}eV^{2}$ and sin$^{2}2\theta$=1.
Since the interesting energy region for atmospheric neutrino oscillation studies
spans from $\simeq$ 1 GeV up to some tens of GeV, it is important to
improve the detector space resolution to push the saturation point 
as high as possible. 
We improved the MACRO space resolution exploiting the TDCs of the MACRO
QTP system\cite{napoletani} to operate the limited streamer tubes in drift mode.
The QTP system is equipped with a 6.6 MHz clock which corresponds to a 
TDC bin size of $\Delta$T=150 ns.
Although the MACRO streamer tubes, operated in drift mode, can reach a
space resolution as good
as $\sigma$$\simeq$250$\mu$m\cite{Battistoni}, 
in MACRO the main limitation comes from the TDC bin size.
The expected ultimate resolution is  
$\sigma$$\simeq$$V_{drift}$$\times$$\Delta$T/$\sqrt{12}$$\simeq$2mm,
where $V_{drift}\simeq$ 4 cm/$\mu$s is the drift velocity.
\EPSFIGURE{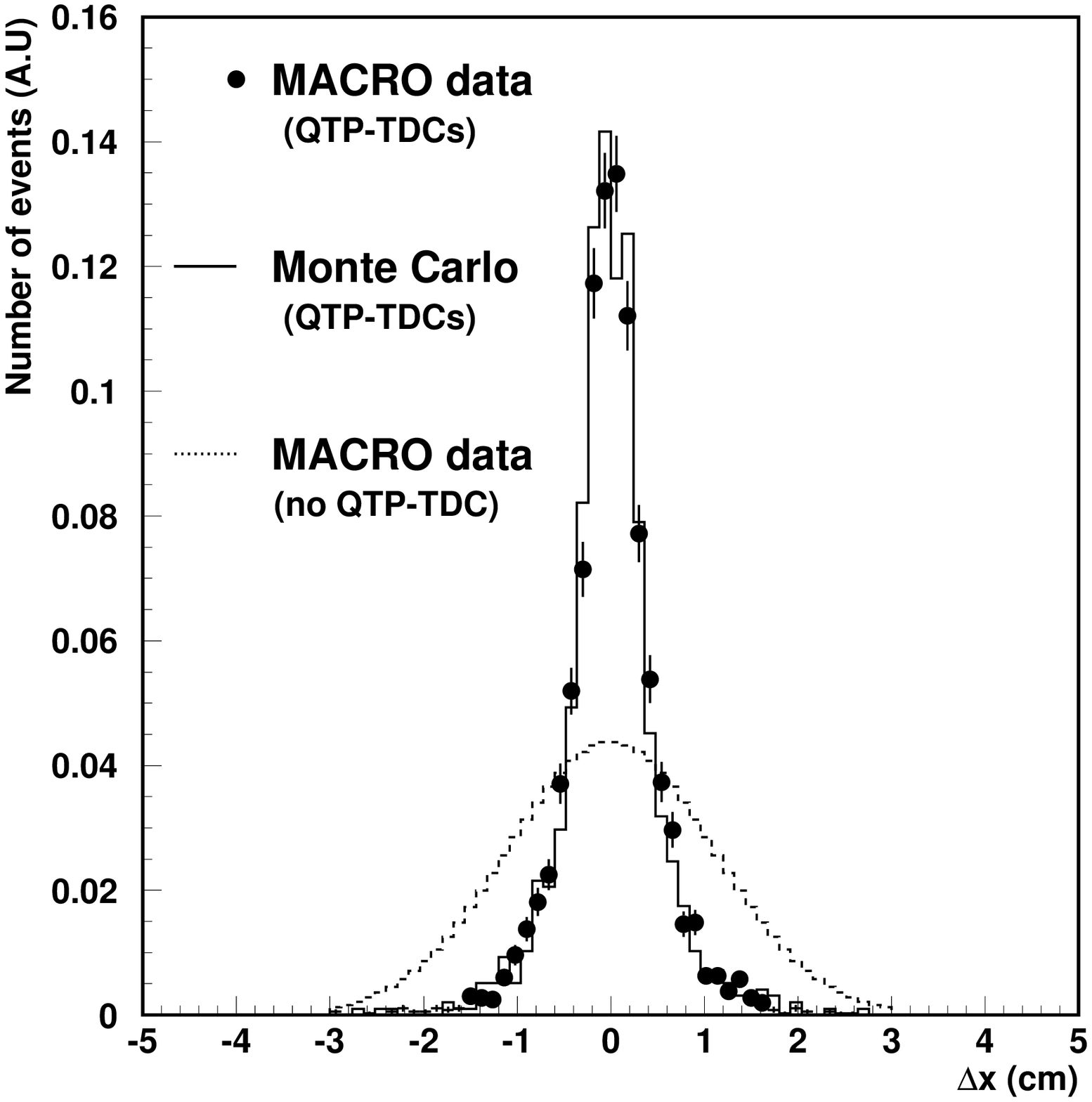,width=7.0cm}
{\it Distribution of the residuals for MACRO data (histogram) 
and for simulated data (black circles). 
The dashed histogram shows the streamer tube resolution used
in digital mode.}
Since the QTP electronics was designed  for slow monopole analysis, 
in order to fully understand the performance of the QTP TDCs in this
context and to perform an absolute energy calibration, we made two tests
at CERN PS-T9 and SPS-X7 beams. A slice of the MACRO detector was
reproduced in detail: absorbers made of rock excavated in the Gran Sasso
tunnel, like those of MACRO, were used. Following the MACRO geometry, the 
tracking was performed by 14 limited streamer tube chambers, 
operated with the MACRO gas mixture (He(73$\%$)/n-pentane(27$\%$)). 
The experimental setup was exposed to
muons with energy ranging from 1 GeV up to 100 GeV.
Each QTP-TDC time was converted into drift circles 
inside the chambers. 
The distribution of the
residuals of the fitted tracks showed a $\sigma \simeq$~2 mm, demonstrating
the successful use of the QTP-TDCs to operate the streamer tube system in
drift mode. 
In order to implement this technique in the MACRO data, 
we used more than 15$\cdot$$10^{6}$ downgoing muons to
align the wire positions with an iterative software procedure. After the
alignment, a resolution of $\sigma \simeq$ 3 mm was obtained. This is 
a factor
3.5 better than the standard resolution of the streamer tube system used in 
digital mode (Fig. 1).
The distribution of the MACRO downgoing muon residuals 
is shown in Fig. 1 (black circles) together with the GMACRO simulation 
(continuous line). In the same plot we superimposed the
residuals distribution obtained with streamer tubes in digital mode
(dashed line).
\EPSFIGURE{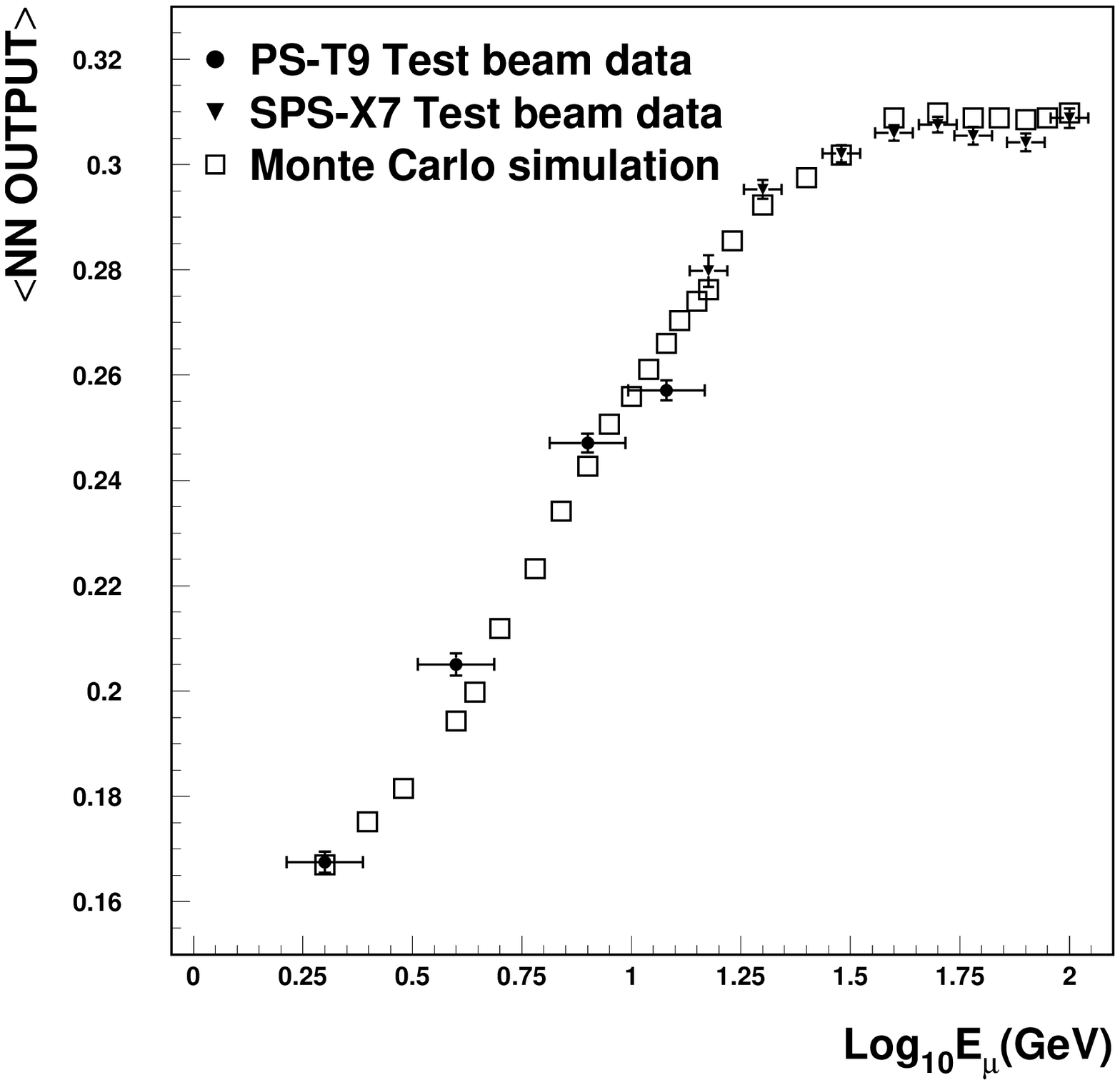,width=9.cm}
{\it Average Neural network output as a function of the muon energy }
The difference between the resolution obtained at test beam ($\sigma
\simeq$ 2 mm) with respect to that obtained with MACRO data ($\sigma
\simeq$ 3 mm), comes from systematic effects such 
as 
the presence of $\delta$~rays
produced in the rock absorbers causing earlier stops to QTP-TDCs. 
\section{Muon energy estimate and data analysis}
For the muon energy estimate we followed a neural network (NN) approach.
We chose JETNET 3.0, a standard package with a multilayer perceptron
architecture and with back-propagation updating. 
The NN has been configured
with 7 input variables, related to the multiple scattering, 1 hidden layer and we chose the Manhattan
upgrading function.
The NN was trained using a set of
Monte Carlo events with known input energy, crossing the detector at
different zenith angles.
In Fig. 2 we show the average output of the NN as a function of the residual 
muon energy before entering the detector.
The output of the NN increases with the muon residual energy up to
$E_{\mu} \simeq$ 40 GeV, ($E_{\nu}\simeq$ 200 GeV).
For the analysis, we used the whole sample of upgoing muon events
collected
with the upper part of MACRO (Attico) running,
for a total live time of 5.5 years. We considered upgoing muons
selected by the TOF system and the muon tracks 
reconstructed with the standard MACRO tracking. We then 
selected hits belonging to the track and made of a single fired tube, 
to  associate unambiguously the QTP-TDC time information. 
Spurious background hits have been avoided by requiring a time window 
of 2 $\mu$s around the trigger time.
Finally, we selected events with at least four streamer tube planes with 
valid QTP-TDC data. 
After the selection cuts 
348 events survived, giving an efficiency of about 50\%.

We used the information provided by the neural network to separate the
upgoing muons into 
different energy regions and to study therein the oscillation
effects. 
We studied the zenith angle
distributions of the upgoing muon events in four regions with different
muon energy, 
selected according to the NN output. 
The same selection has been applied to simulated events.
To make a comparison between real data and Monte Carlo
expectations, we performed a full simulation chain by using the Bartol 
neutrino flux and the GRV94 DIS parton distributions
\cite{Gluck}.
\EPSFIGURE{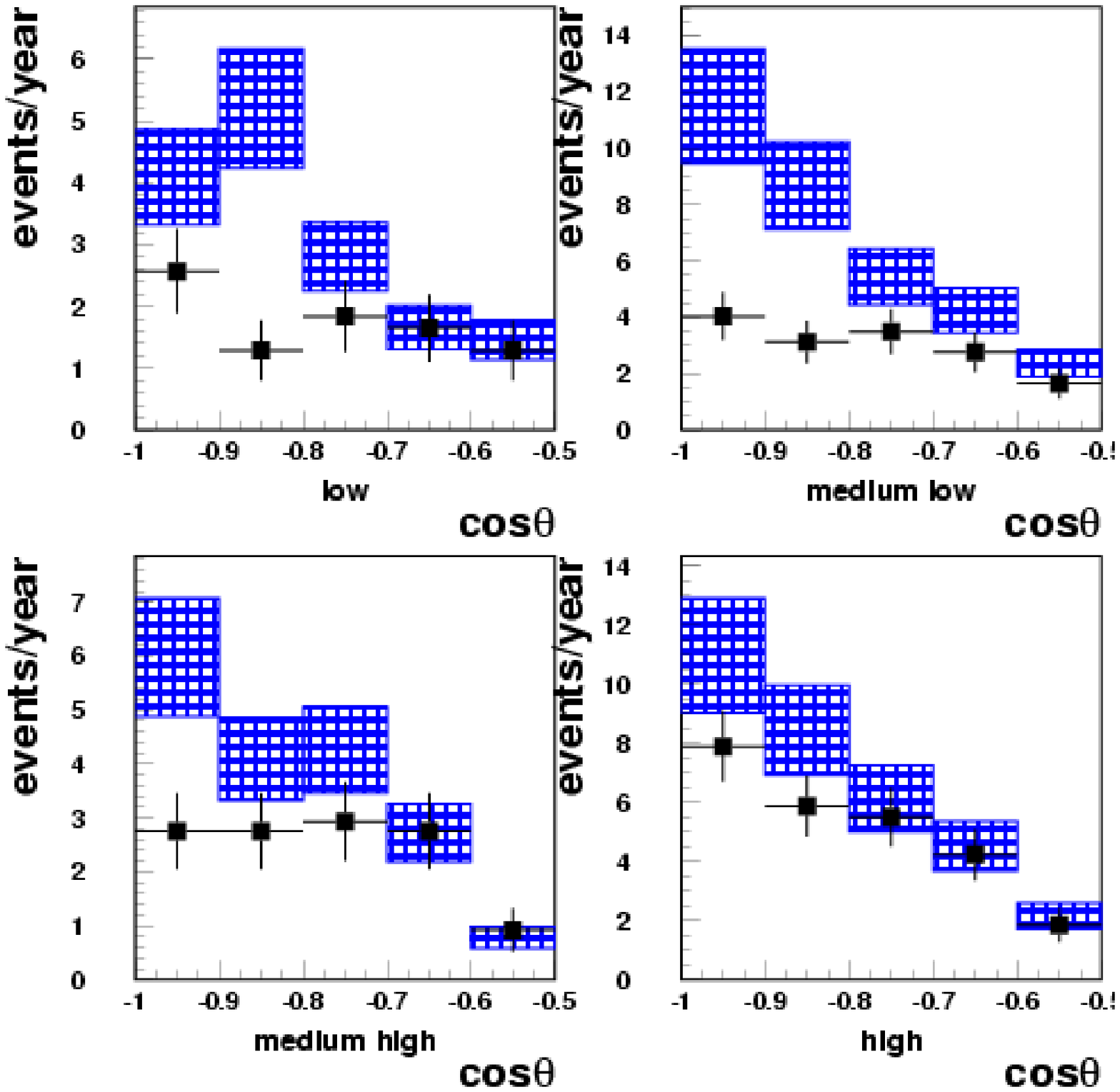,width=10cm}
{\it Zenith angle distributions for upward going muons in four energy
    windows (black squares). Rectangular boxes show the Monte Carlo expectation
    with the no-oscillation hypothesis (statistical errors plus 17\%
    systematic uncertainty on the overall flux).}
The propagation of the muons from the interaction point up to the
detector has been done using the FLUKA99 package\cite{fluka},
while the muon simulation inside the detector was performed with GMACRO 
(the GEANT 3.21 based detector simulation).

Should the upgoing muon deficit and the angular distribution distortion 
(with respect to the Monte Carlo expectation) pointed out
by MACRO come from neutrino oscillations with parameters 
$\Delta m^{2}$=$\cal{O}$($10^{-3}$ $eV^{2}$) and sin$^{2}$2$\theta$$\simeq$1,
such deficit and such angular distribution distortion would not manifest
at all neutrino energies. The effect is expected to be stronger at low
neutrino energies (E$\leq$ 10 GeV) and to disappear at higher energies
(E$\geq$100 GeV).
We used the NN to separate four different neutrino energy regions whose median 
energy is respectively 12 GeV (low), 20 GeV (medium-low), 
50 GeV (medium high) and 102 GeV (high).
In Fig. 3 we show the zenith angle distributions of the upgoing muon events in
the four energy regions selected compared to the expectations of Monte Carlo
simulation, assuming the no-oscillation hypothesis.
It is evident that at low energy a strong disagreement
between data and Monte Carlo (no-oscillation hypothesis) is present,
while the agreement is restored with increasing  neutrino energy.
The corresponding $\chi^{2}$-probabilities for the no-oscillation 
hypothesis in these four windows are respectively 
1.8$\%$ (low), 16.8$\%$ (medium-low), 26.9$\%$ (medium-high) and 
87.7$\%$ (high): 
the $\chi^{2}$/DoF values are clearly running with the neutrino energy,
spanning from 13.7/5 to 1.8/5.
The $\chi^{2}$ has been computed using only the angular shape.
Finally, we tried to get information on the ratio $L_{\nu}/E_{\nu}$.
The output of the NN was calibrated on an {\it event by event} basis to
have a linear response as a function of $log_{10}$($L_{\nu}/E_{\nu}$).
The  ratio of  DATA/ Monte Carlo( no oscillation) as a function 
of $log_{10}(L_{\nu}/E_{\nu})$,
is plotted in Fig. 4: a good agreement is found with the
oscillation probability function we expect with the parameters quoted above.

\section{Conclusions}
The sample of upward through-going muons measured by MACRO has been
analyzed in terms of neutrino oscillations using 
multiple Coulomb scattering to infer muon energy. 
The improvement of the space resolution 
obtained by exploiting the QTP electronics
extended the muon residual energy reconstruction 
up to $\simeq$ 40 GeV. Two dedicated runs at the
\EPSFIGURE{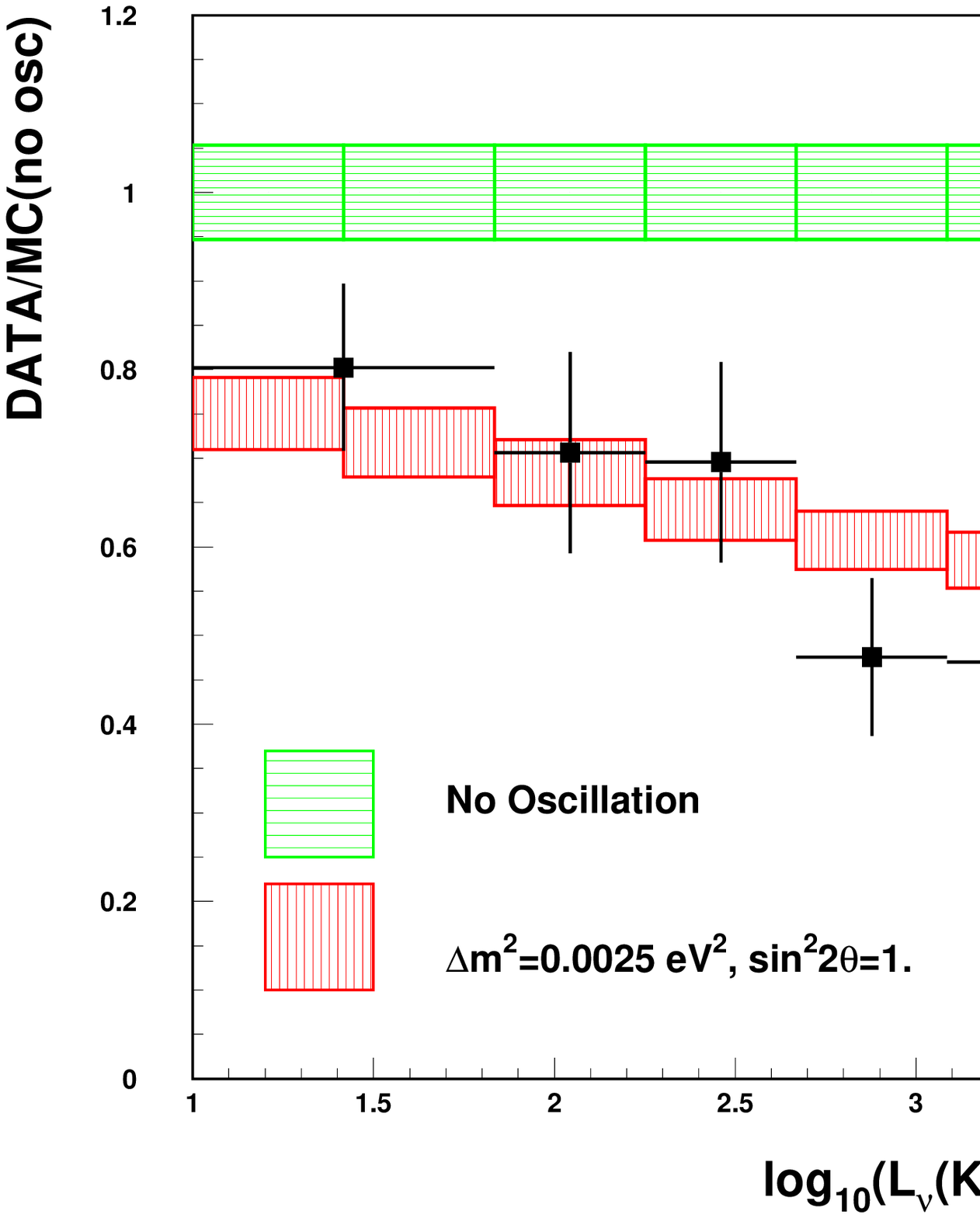,width=9cm}
{\it Data/MC (without oscillation) as a function of the ratio
$L_{\nu}/E_{\nu}$.}
CERN PS-T9 and SPS-X7 beams allowed us to check the MACRO QTP-TDCs and showed the
feasibility of operating the limited streamer tubes in drift mode. 
The angular distribution of the upward going muon sample has
been subdivided into four energy windows, 
showing the energy trend expected from
the neutrino oscillation hypothesis. Moreover, we performed a study in terms 
of $L_{\nu}/E_{\nu}$. Also in this case, the observed transition from 1 to
0.5 in the ratio of data to Monte Carlo prediction 
is the one expected from the neutrino oscillation hypothesis 
with oscillation parameters 
$\Delta m^{2}=$ $\cal{O}$($10^{-3}$ $eV^{2}$) and sin$^{2}2\theta$=1.



\begin{thebibliography}{99}
\bibitem{Ahlen} 
  S. Ahlen et al, The MACRO Coll., Nucl. Instrum. Meth A324(1993)337.
\bibitem{macro}
  S. Ahlen et al, The MACRO Coll., Phys. Lett. B357(1995)481;
  M. Ambrosio et al., MACRO Coll., Phys. Lett. B434(1998)451.
\bibitem{napoletani} M. Ambrosio et al.,  Nucl. Instrum. \& Meth A321(1992)609.
\bibitem{Battistoni} 
  G. Battistoni et al., [hep-ph/0105099] , Accepted by
Nucl. Instrum. Meth. A.
\bibitem{Bakari}
  D. Bakari et al, for the MACRO Coll., Talk given at Advanced NATO 
Workshop, 21-23 March 2001, Ojuda(Marocco),[hep-ex/0105087].
\bibitem{bartol}
  V. Agrawal et al., Phys. Rev. D53(1996)1314.
\bibitem{Gluck}
  M. Gluck et al, Z. Phys. C67(1995)433.
\bibitem{fluka} 
  A. Fasso', et al., Proc. 2nd workshop on Simulating Accelerator Radiation
  Environment, SARE-2, CERN-Geneva, 9-11, October, 1995.
\end{thebibliography}
\end{document}